\def\tr#1{{\rm tr} #1}

\documentclass{livrev}
\begin{document}

\title{Theorems on Existence and Global Dynamics for the Einstein Equations}
\author{Alan D.\ Rendall\\
        Max-Planck-Institut f{\"u}r Gravitationsphysik\\
        Am M{\"u}hlenberg 1, 14424 Golm, Germany\\
        rendall@aei-potsdam.mpg.de\\
        http://www.aei-potsdam.mpg.de/\~{}rendall/rendall.html}

\date{}
\maketitle

\begin{abstract}
This article is a guide to theorems on existence and global
dynamics of solutions of the Einstein equations. It draws attention to 
open questions in the field. The local in time Cauchy problem, which is
relatively well understood, is surveyed. Global results for solutions 
with various types of symmetry are discussed. A selection of results 
from Newtonian theory and special relativity which offer useful comparisons
is presented. Treatments of global results in the case of small data and 
results on constructing spacetimes with prescribed singularity structure
are given. A conjectural picture of the asymptotic behaviour of general
cosmological solutions of the Einstein equations is built up. Some 
miscellaneous topics connected with the main theme are collected in a
separate section.
\end{abstract}


\newpage
\section{Introduction}\label{introduction}

Systems of partial differential equations are of central importance in 
physics. Only the simplest of these equations can be solved by explicit 
formulae. Those which cannot are commonly studied by means of 
approximations. There is, however, another approach which
is complementary. This consists in determining the qualitative
behaviour of solutions, without knowing them explicitly. The first
step in doing this is to establish the existence of solutions under 
appropriate circumstances. Unfortunately, this is often hard, and 
obstructs the way to obtaining more interesting information. When 
partial differential equations are investigated with a view to 
applications existence theorems should not become a goal in themselves.
It is important to remember that, from a more general point of view, they
are only a starting point.

The basic partial differential equations of general relativity are
Einstein's equations. In general they are coupled to other partial
differential equations describing the matter content of spacetime.
The Einstein equations are essentially hyperbolic in nature. In
other words, the general properties of solutions are similar to those
found for the wave equation. It follows that it is reasonable to try to
determine a solution by initial data on a spacelike hypersurface.
Thus the Cauchy problem is the natural context for existence theorems
for the Einstein equations. The Einstein equations are also nonlinear.
This means that there is a big difference between the local and global 
Cauchy problems. A solution evolving from regular data may develop 
singularities.

A special feature of the Einstein equations is that they are
diffeomorphism invariant. If the equations are written down in
an arbitrary coordinate system then the solutions of these coordinate
equations are not uniquely determined by initial data. Applying a
diffeomorphism to one solution gives another solution. If this 
diffeomorphism is the identity on the chosen Cauchy surface up to
first order then the data are left unchanged by this transformation.
In order to obtain a system for which uniqueness in the Cauchy problem
holds in the straightforward sense it does for the wave equation, some 
coordinate or gauge fixing must be carried out.

Another special feature of the Einstein equations is that initial 
data cannot be prescribed freely. They must satisfy constraint
equations. To prove the existence of a solution of the Einstein
equations, it is first necessary to prove the existence of a
solution of the constraints. The usual method of solving the
constraints relies on the theory of elliptic equations.

The local existence theory of solutions of the Einstein equations
is rather well understood. Section~\ref{local} points out some of the things 
which are not known. On the other hand, the problem of proving general 
global existence theorems for the Einstein equations is beyond the reach of 
the mathematics presently available. To make some progress, it is necessary 
to concentrate on simplified models. The most common simplifications are
to look at solutions with various types of symmetry and solutions for
small data. These two approaches are reviewed in sections~\ref{symmetric}
and~\ref{small} respectively. A different approach is to prove the 
existence of solutions with a prescribed singularity structure. This is
discussed in section~\ref{prescribe}. Section~\ref{further} collects some 
miscellaneous results which cannot easily be classified. With the motivation
that insights about the properties of solutions of the Einstein equations
can be obtained from the comparison with Newtonian theory and special
relativity, relevant results from those areas are presented in section
\ref{newtonian}.

The sections just listed are to some extent catalogues of known results,
augmented with some suggestions as to how these could be extended in the
future. Sections \ref{expand} and \ref{sing} complement this by looking  
ahead to see what the final answer to some interesting general questions
might be. They are necessarily more speculative than the other sections
but are rooted in the known results surveyed elsewhere in the article. 

The area of research reviewed in the following relies heavily on the theory
of differential equations, particularly that of hyperbolic partial 
differential equations. For the benefit of readers with little background
in differential equations, some general references which the author has
found to be useful will be listed. A thorough introduction to ordinary
differential equations is given in~\cite{hartman82}. A lot of intuition
for ordinary differential equations can be obtained from~\cite{hubbard91}.
The article~\cite{arnold88} is full of information, in rather compressed
form. A classic introductory text on partial differential equations, where
hyperbolic equations are well represented, is~\cite{john82}. Useful texts
on hyperbolic equations, some of which explicitly deal with the Einstein
equations, are~\cite{taylor96,kichenassamy96a,racke92,majda84,strauss89,john90,
evans98}. 

An important aspect of existence theorems in general relativity which
one should be aware of is their relation to the cosmic censorship 
hypothesis. This point of view was introduced in an influential paper by
Moncrief and Eardley~\cite{moncrief81a}. An extended discussion of the
idea can be found in~\cite{chrusciel91a}.  

\section{Local existence}\label{local}

In this section basic facts about local existence theorems for the Einstein
equations are recalled. Since the theory is well developed and good 
accounts exist elsewhere (see for instance \cite{friedrich00a}), attention 
is focussed on remaining open questions known to the author. In particular, 
the questions of the minimal regularity required to get a well-posed problem 
and of free boundaries for fluid bodies are discussed.

\subsection{The constraints}\label{constraints}

The unknowns in the constraint equations are the initial data for the 
Einstein equations. These consist of a three-dimensional manifold $S$,
a Riemannian metric $h_{ab}$ and a symmetric tensor $k_{ab}$ on $S$,
and initial data for any matter fields present. The equations are: 
\begin{equation}
R-k_{ab}k^{ab}+(h^{ab}k_{ab})^2=16\pi\rho
\label{equation1}
\end{equation}
\begin{equation}         \\ 
\nabla^a k_{ab}-\nabla_b(h^{ac}k_{ac})=8\pi j_b   
\label{equation2}
\end{equation}
Here $R$ is the scalar curvature of the metric $h_{ab}$ and $\rho$ and $j_a$
are projections of the energy-momentum tensor. Assuming matter fields which
satisfy the dominant energy condition implies that $\rho\ge (j_aj^a)^{1/2}$.
This means that the trivial procedure of making an arbitrary choice of 
$h_{ab}$ and $k_{ab}$ and defining $\rho$ and $j_a$ by equations~(\ref{equation1}) and~(\ref{equation2}) 
is of no use for producing physically interesting solutions. 

The usual method for solving the equations~(\ref{equation1}) and~(\ref{equation2}) is the conformal 
method~\cite{choquet80}. In this
method parts of the data (the so-called free data) are chosen, and the
constraints imply four elliptic equations for the remaining parts. The
case which has been studied most is the constant mean curvature (CMC)
case, where $\tr k=h^{ab}k_{ab}$ is constant. In that case there is an
important simplification. Three of the elliptic equations, which form
a linear system, decouple from the remaining one. This last equation,
which is nonlinear, but scalar, is called the Lichnerowicz equation.
The heart of the existence theory for the constraints in the CMC case
is the theory of the Lichnerowicz equation.

Solving an elliptic equation is a non-local problem and so boundary
conditions or asymptotic conditions are important. For the constraints
the cases most frequently considered in the literature are that where
$S$ is compact (so that no boundary conditions are needed) and that
where the free data satisfy some asymptotic flatness conditions. In
the CMC case the problem is well understood for both kinds of boundary 
conditions~\cite{cantor79,christodoulou81,isenberg95}. The other case which 
has been studied in detail is that of hyperboloidal data~\cite{andersson92}. 
The kind of theorem which is obtained is that sufficiently differentiable 
free data, in some cases required to satisfy some global restrictions, can be 
completed in a unique way to a solution of the constraints. It should be
noted in passing that in certain cases physically interesting free data
may not be \lq sufficiently differentiable\rq\ in the sense it is meant
here. One such case is mentioned at the end of section \ref{freeboundary}.
The usual kinds of differentiability conditions which are required in the
study of the constraints involve the free data belonging to suitable
Sobolev or H\"older spaces. Sobolev spaces have the advantage that they
fit well with the theory of the evolution equations. (Compare the discussion
in section \ref{vacuum}.) In the literature nobody seems to have focussed on 
the question of the minimal differentiability necessary in order to apply 
the conformal method. 

In the non-CMC case our understanding is much more limited although
some results have been obtained in recent years (see~\cite{isenberg96,
choquet00a} and references therein.) It is an 
important open problem to extend these so that an overview is obtained
comparable to that available in the CMC case. Progress on this could 
also lead to a better understanding of the question, when a spacetime
which admits a compact, or asymptotically flat, Cauchy surface also
admits one of constant mean curvature. Up to now there are only
isolated examples which exhibit obstructions to the existence of CMC
hypersurfaces~\cite{bartnik88a}.

It would be interesting to know whether there is a useful concept of
the most general physically reasonable solutions of the constraints
representing regular initial configurations. Data of this kind should not 
themselves contain singularities. Thus it seems reasonable to suppose at
least that the metric $h_{ab}$ is complete and that the length of $k_{ab}$, 
as measured using $h_{ab}$, is bounded. Does the existence of solutions of 
the constraints imply a restriction on the topology of $S$ or on the 
asymptotic geometry of the data? This question is largely open, and it seems
that information is available only in the compact and asymptotically flat 
cases. In the case of compact $S$, where there is no asymptotic regime,
there is known to be no topological restriction. In the asymptotically flat 
case there is also no topological restriction implied by the constraints
beyond that implied by the condition of asymptotic flatness itself~\cite{witt86}
This shows in particular that any manifold which is obtained by deleting a
point from a compact manifold admits a solution of the constraints satisfying
the minimal conditions demanded above. A starting point for going beyond
this could be the study of data which are asymptotically homogeneous. For 
instance, the Schwarzschild solution contains interesting CMC hypersurfaces 
which are asymptotic to the metric product of a round 2-sphere with the real 
line. More general data of this kind could be useful for the study of the 
dynamics of black hole interiors~\cite{rendall96a}.

To sum up, the conformal approach to solving the constraints, which is
the standard one up to now, is well understood in the compact, asymptotically 
flat and hyperboloidal cases under the constant mean curvature assumption,
and only in these cases. For some other approaches see~\cite{bartnik93a},
~\cite{bartnik93b} and~\cite{york99}. New techniques have been applied by
Corvino \cite{corvino00a} to prove the existence of regular solutions of 
the vacuum constraints on ${\bf R}^3$ which are Schwarzschild outside a 
compact set.

\subsection{The vacuum evolution equations}\label{vacuum}

The main aspects of the local in time existence theory for the Einstein
equations can be illustrated by restricting to smooth (i.~e.\ infinitely 
differentiable) data for the vacuum Einstein equations. The generalizations
to less smooth data and matter fields are discussed in sections  
\ref{differentiability} and~\ref{matter} respectively. In the vacuum 
case the data are $h_{ab}$ and $k_{ab}$ on a three-dimensional manifold $S$, 
as discussed in section~\ref{constraints}.  
A solution corresponding to these data is given by a 
four-dimensional manifold $M$, a Lorentz metric $g_{\alpha\beta}$ on
$M$ and an embedding of $S$ in $M$. Here $g_{\alpha\beta}$ is supposed to
be a solution of the vacuum Einstein equations while $h_{ab}$ and $k_{ab}$
are the induced metric and second fundamental form of the embedding,
respectively.

The basic local existence theorem says that, given smooth data for the 
vacuum Einstein equations, there exists a smooth solution of the equations
which gives rise to these data~\cite{choquet80}. Moreover, it can be assumed 
that the image of $S$ under the given embedding is a Cauchy surface for the 
metric $g_{\alpha\beta}$. The latter fact may be expressed loosely, 
identifying $S$ with its image, by the statement that $S$ is a Cauchy 
surface. A solution of the Einstein equations with given initial data having 
$S$ as a Cauchy surface is called a Cauchy development of those data. The 
existence theorem is local because it says nothing about the size of the 
solution obtained. A Cauchy development of given data has many open subsets 
which are also Cauchy developments of that data.

It is intuitively clear what it means for one Cauchy development to be
an extension of another. The extension is called proper if it is strictly
larger than the other development. A Cauchy development which has no proper
extension is called maximal. The standard global uniqueness theorem for
the Einstein equations uses the notion of the maximal development. It is
due to Choquet-Bruhat and Geroch~\cite{choquet69}. It says that the maximal 
development of any Cauchy data is unique up to a diffeomorphism which fixes 
the initial hypersurface. It is also possible to make a statement of Cauchy 
stability which says that, in an appropriate sense, the solution depends 
continuously on the initial data. Details on this can be found in
~\cite{choquet80}.

A somewhat stronger form of the local existence theorem is to say that
the solution exists on a uniform time interval in all of space. The 
meaning of this is not a priori clear, due to the lack of a preferred 
time coordinate in general relativity. The following is a formulation which 
is independent of coordinates. Let $p$ be a point of $S$. The temporal
extent $T(p)$ of a development of data on $S$ is the supremum of the length 
of all causal curves in the development passing through $p$. In this way a
development defines a function $T$ on $S$. The development can be regarded
as a solution which exists on a uniform time interval if $T$ is bounded
below by a strictly positive constant. For compact $S$ this is a 
straightforward consequence of Cauchy stability. In the case of asymptotically 
flat data it is less trivial. In the case of the vacuum Einstein equations it 
is true, and in fact the function $T$ grows at least linearly as function
of spatial distance at infinity ~\cite{christodoulou81}. It should follow
from the results of \cite{klainerman99} that the constant of proportionality
in the linear lower bound for $T$ can be chosen to be unity, but this does 
not seem to have been worked out explicitly.   

When proving the above local existence and global uniqueness theorems it
is necessary to use some coordinate or gauge conditions. At least no
explicitly diffeomorphism-invariant proofs have been found up to now.
Introducing these extra elements leads to a system of reduced equations,
whose solutions are determined uniquely by initial data in the strict 
sense, and not just uniquely up to diffeomorphisms. When a solution of the
reduced equations has been obtained, it must be checked that it is a 
solution of the original equations. This means checking that the constraints
and gauge conditions propagate. There are many methods for reducing the
equations. An overview of the possibilities may be found 
in~\cite{friedrich96}. See also \cite{friedrich00a}.

\subsection{Questions of differentiability}\label{differentiability}

Solving the Cauchy problem for a system of partial differential equations
involves specifying a set of initial data to be considered, and determining
the differentiability properties of solutions. Thus two regularity properties
are involved -- the differentiability of the allowed data, and that of the
corresponding solutions. Normally it is stated that for all data 
with a given regularity, solutions with a certain type of regularity
are obtained. For instance in the section~\ref{vacuum} we chose both types of
regularity to be `infinitely differentiable'. The correspondence
between the regularity of data and that of solutions is not a matter of
free choice. It is determined by the equations themselves, and in
general the possibilities are severely limited. A similar issue arises
in the context of the Einstein constraints, where there is a correspondence
between the regularity of free data and full data.

The kinds of regularity properties which can be dealt with in the Cauchy
problem depends of course on the mathematical techniques available.
When solving the Cauchy problem for the Einstein equations it is necessary
to deal at least with nonlinear systems of hyperbolic equations. (There
may be other types of equations involved, but they will be ignored here.)
For general nonlinear systems of hyperbolic equations the standard
technique is the method of energy estimates. This method is
closely connected with Sobolev spaces, which will now be discussed briefly.

Let $u$ be a real-valued function on ${\bf R}^n$. Let:
$$\|u\|_s=(\sum_{i=0}^s\int |D^i u|^2 (x) dx)^{1/2}.$$
The space of functions for which this quantity is finite is the Sobolev
space $H^s({\bf R}^n)$. Here $|D^i u|^2$ denotes the sum of the squares of 
all partial derivatives of $u$ of order $i$. Thus the Sobolev space
$H^s$ is the space of functions, all of whose partial derivatives up
to order $s$ are square integrable. Similar spaces can be defined for
vector valued functions by taking a sum of contributions from the
separate components in the integral. It is also possible to define 
Sobolev spaces on any Riemannian manifold, using covariant derivatives.
General information on this can be found in~\cite{aubin82}. Consider now a
solution $u$ of the wave equation in Minkowski space. Let $u(t)$ be
the restriction of this function to a time slice. Then it is easy to
compute that, provided $u$ is smooth and $u(t)$ has compact support for 
each $t$, the quantity $\|Du(t)\|^2_s+\|\partial_t u(t)\|^2_s$ is time 
independent for each $s$. For $s=0$ this is just the energy of a solution of 
the wave equation. For a general nonlinear hyperbolic system, the Sobolev 
norms are no longer time-independent. The constancy in time is replaced by 
certain inequalities. Due to the similarity to the energy for the wave 
equation, these are called energy estimates. They constitute the foundation 
of the theory of hyperbolic equations. It is because of these estimates that
Sobolev spaces are natural spaces of initial data in the Cauchy problem
for hyperbolic equations. The energy estimates ensure that a solution
evolving from data belonging to a given Sobolev space on one spacelike
hypersurface will induce data belonging to the same Sobolev space on later
spacelike hypersurfaces. In other words, the property of belonging to a
Sobolev space is propagated by the equations. Due to the locality properties 
of hyperbolic equations (existence of a finite domain of dependence), it is 
useful to introduce the spaces $H^s_{\rm loc}$ which are defined by the 
condition that whenever the domain of integration is restricted to a compact 
set the integral defining the space $H^s$ is finite.

In the end the solution of the Cauchy problem should be a function
which is differentiable enough in order that all derivatives 
which occur in the equation exist in the usual (pointwise) sense.
A square integrable function is in general defined only almost everywhere
and the derivatives in the above formula must be interpreted as
distributional derivatives. For this reason a connection between Sobolev 
spaces and functions whose derivatives exist pointwise is required. This is 
provided by the Sobolev embedding theorem. This says that if a function $u$ 
on ${\bf R}^n$ belongs to the Sobolev space $H^s_{\rm loc}$ and if $k<s-n/2$ 
then there is a $k$ times continuously differentiable function which agrees 
with $u$ except on a set of measure zero.

In the existence and uniqueness theorems stated in section~\ref{vacuum}, the
assumptions on the initial data for the vacuum Einstein equations can be 
weakened to say that $h_{ab}$ should belong to $H^s_{\rm loc}$ and $k_{ab}$ 
to $H^{s-1}_{\rm loc}$. Then, provided $s$ is large enough, a solution is 
obtained which belongs to $H^s_{\rm loc}$. In fact its restriction to
any spacelike hypersurface also belongs to $H^s_{\rm loc}$, a property
which is a priori stronger. The details of how large $s$ must be would be out 
of place here, since they involve examining the detailed structure of the 
energy estimates. However there is a simple rule for computing the required 
value of $s$. The value of $s$ needed to obtain an existence theorem for the 
Einstein equations using energy estimates is that for which the Sobolev 
embedding theorem, applied to
spatial slices, just ensures that the metric is continuously differentiable.
Thus the requirement is that $s>n/2+1=5/2$, since $n=3$. It follows that the 
smallest possible integer $s$ is three. Strangely enough, uniqueness up to 
diffeomorphisms is only known to hold for $s\ge 4$. The reason is that in 
proving the uniqueness theorem a diffeomorphism must be carried out, which 
need not be smooth. This apparently leads to a loss of one derivative.
It would be desirable to show that uniqueness holds for $s=3$ and to
close this gap, which has existed for many years. There exists a definition
of Sobolev spaces for an arbitrary real number $s$, and hyperbolic 
equations can also be solved in the spaces with $s$ not an integer 
~\cite{taylor91}. Presumably these techniques could be applied to prove local 
existence for the Einstein equations with $s$ any real number greater than 
$5/2$. However this has apparently not been done explicitly in the literature.

Consider now $C^\infty$ initial data. Corresponding to these data there is
a development of class $H^s$ for each $s$. It could conceivably be the case
that the size of these developments shrinks with increasing $s$. In that 
case their intersection might contain no open neighbourhood of the initial
hypersurface, and no smooth development would be obtained. Fortunately it
is known that the $H^s$ developments cannot shrink with increasing $s$,
and so the existence of a $C^\infty$ solution is obtained for $C^\infty$
data. It appears that the $H^s$ spaces with $s$ sufficiently large are the 
only spaces containing the space of smooth functions for which it has been 
proved that the Einstein  equations are locally solvable.

What is the motivation for considering regularity conditions other than
the apparently very natural $C^\infty$ condition?  One motivation concerns
matter fields and will be discussed in section~\ref{matter}. Another is 
the idea that assuming the existence of many derivatives which have no direct
physical significance seems like an admission that the problem has not
been fully understood. A further reason for considering low regularity
solutions is connected to the possibility of extending a local existence
result to a global one. If the proof of a local existence theorem is
examined closely it is generally possible to give a continuation criterion.
This is a statement that if a solution on a finite time interval is such 
that a certain quantity constructed from the solution is bounded on that 
interval, then the solution can be extended to a longer time interval. (In
applying this to the Einstein equations we need to worry about introducing 
an appropriate time coordinate.) If it can be shown that the relevant 
quantity is bounded on any finite time interval where a solution exists, 
then global existence follows. It suffices to consider the maximal interval 
on which a solution is defined, and obtain a contradiction if that interval
is finite. This description is a little vague, but contains the essence of
a type of argument which is often used in global existence proofs. The
problem in putting it into practice is that often the quantity whose
boundedness has to be checked contains many derivatives, and is therefore
difficult to control. If the continuation criterion can be improved by
reducing the number of derivatives required, then this can be a significant
step towards a global result. Reducing the number of derivatives in the
continuation criterion is closely related to reducing the number of 
derivatives of the data required for a local existence proof.

A striking example is provided by the work of Klainerman and Machedon
~\cite{klainerman95} on the Yang-Mills equations in Minkowski space. Global 
existence in this case was first proved by Eardley and Moncrief
~\cite{eardley82}, assuming initial data of
sufficiently high differentiability. Klainerman and Machedon gave a new 
proof of this which, though technically complicated, is based on a
conceptually simple idea. They prove a local existence theorem for data 
of finite energy. Since energy is conserved this immediately proves 
global existence. In this case finite energy corresponds to the Sobolev
space $H^1$ for the gauge potential. Of course a result of this kind 
cannot be expected for the Einstein equations, since spacetime singularities
do sometimes develop from regular initial data. However, some weaker 
analogue of the result could exist. 

\subsection{New techniques for rough solutions}\label{rough}

Recently new mathematical techniques have been developed to lower the 
threshold of differentiability required to obtain local existence for
quasilinear wave equations in general and the Einstein equations in
particular. Some aspects of this development will now be discussed 
following \cite{klainerman01a} and \cite{klainerman01b}. A central
aspect is that of Strichartz inequalities. These allow one to go
beyond the theory based on $L^2$ spaces and use Sobolev spaces based
on the Lebesgue $L^p$ spaces for $p\ne 2$. The classical approach to
deriving Strichartz estimates is based on the Fourier transform and
applies to flat space. The new ideas allow the use of the Fourier
transform to be limited to that of Littlewood-Paley theory and 
facilitate generalizations to curved space.

The idea of Littlewood-Paley theory is as follows. (See \cite{alinhac91}
for a good exposition of this.) Suppose that we want to describe the
regularity of a function (or, more generally, a tempered distribution)
$u$ on ${\bf R}^n$. Differentiability properties of $u$ correspond, roughly 
speaking, to fall-off properties of its Fourier transform $\hat u$. This
is because the Fourier transform converts differentiation into 
multiplication. The Fourier transform is decomposed as $\hat u=\sum\phi_i u$,
where $\phi_i$ is a dyadic partition of unity. The statement that it is 
dyadic means that all the $\phi_i$ except one are obtained from each other
by scaling the argument by a factor which is a power of two. Transforming
back we get the decomposition $u=\sum u_i$, where $u_i$ is the inverse Fourier
transform of $\phi_i u$. The component $u_i$ of $u$ contains only frequencies
of the order $2^i$. In studying rough solutions of the Einstein equations
the Littlewood-Paley decomposition is applied to the metric itself. The 
high frequencies are discarded to obtain a smoothed metric which plays an
important role in the arguments.

Another important element of the proofs is to rescale the solution by a factor
depending on the cut-off $\lambda$ applied in the Littlewood-Paley
decomposition. Proving the desired estimates then comes down to proving
the existence of the rescaled solutions on a time interval depending on
$\lambda$ in a particular way. The rescaled data are small in some sense
and so a connection is established to the question of long-time existence
of solutions of the Einstein equation for small initial data. In this way
techniques from the work of Christodoulou and Klainerman on the stability
of Minkowski space (see section \ref{minkowski}) are brought in.

What is finally proved? In general there is a close connection between
proving local existence for data in a certain space and showing that the 
time of existence of smooth solutions only depends on the norm of the
data in the given space. Klainerman and Rodnianski \cite{klainerman01b} 
demonstrate that the time of existence of solutions of the reduced
Einstein equations in harmonic coordinates depends only on the 
$H^{2+\epsilon}$ norm of the initial data for any $\epsilon>0$. 
The reason that this does not allow them to assert an existence result in 
the same space is that the constraints are needed in their proof and that
an understanding of solving the constraints at this low level of 
differentiability is lacking.

The techniques discussed in this section which have been stimulated by the
desire to understand the Einstein equations are also helpful in understanding
other nonlinear wave equations. Thus this is an example where information
can flow from general relativity to the theory of partial differential
equations.

\subsection{Matter fields}\label{matter}
Analogues of the results for the vacuum Einstein equations given in section
\ref{vacuum}
are known for the Einstein equations coupled to many types of matter
model. These include perfect fluids, elasticity theory, kinetic theory,
scalar fields, Maxwell fields, Yang-Mills fields and combinations of these.
An important restriction is that the general results for perfect fluids
and elasticity apply only to situations where the energy density is 
uniformly bounded away from zero on the region of interest. In particular
they do not apply to cases representing material bodies surrounded by 
vacuum. In cases where the energy density, while everywhere positive,
tends to zero at infinity, a local solution is known to exist, but it is
not clear whether a local existence theorem can be obtained which is 
uniform in time. In cases where the fluid has a sharp boundary,
ignoring the boundary leads to solutions of the Einstein-Euler equations 
with low differentiability (cf.\ section~\ref{differentiability}), while taking
it into account explicitly leads to a free boundary problem. This will be 
discussed in more detail in section~\ref{freeboundary}.
In the case of kinetic or field theoretic matter models it makes no 
difference whether the energy density vanishes somewhere or not.

\subsection{Free boundary problems}\label{freeboundary}

In applying general relativity one would like to have solutions of the
Einstein-matter equations modelling material bodies. As will be discussed
in section~\ref{stationary} there are solutions available for describing
equilibrium situations. However dynamical situations require solving a free
boundary problem if the body is to be made of fluid or an elastic solid.
We will now discuss the few results which are known on this subject. For
a spherically symmetric self-gravitating fluid body in general relativity
a local in time existence theorem was proved in~\cite{kind93}. This 
concerned the case where the density of the fluid at the boundary is
non-zero. In~\cite{rendall92b} a local existence theorem was proved for certain
equations of state with vanishing boundary density. These solutions need
not have any symmetry but they are very special in other ways. In particular
they do not include small perturbations of the stationary solutions
discussed in section~\ref{stationary}. There is no general result on this
problem up to now.

Remarkably, the free boundary problem for a fluid body is also poorly
understood in classical physics. There is a result for a viscous fluid
~\cite{secchi91} but in the case of a perfect fluid the problem was wide 
open until very recently. Now a major step forward has been taken by 
Wu~\cite{wu99}, who obtained a result for a fluid which is incompressible
and irrotational. There is a good physical reason why local existence for
a fluid with a free boundary might fail. This is the Rayleigh-Taylor
instability which involves perturbations of fluid interfaces which grow
with unbounded exponential rates. (Cf.\ the discussion in~\cite{beale93}.)
It turns out that in the case considered by Wu this instability does not 
cause problems and there is no reason to expect that a self-gravitating
compressible fluid with rotation in general relativity with a free 
boundary cannot also be described by a well-posed free boundary value 
problem. For the generalization of the problem considered by Wu to the
case of a fluid with rotation Christodoulou and Lindblad 
\cite{christodoulou00a} have obtained estimates which look as if they
should be enough to obtain an existence theorem. It has, however, not
yet been possible to complete the argument. This point deserves some 
further comment. In many problems the heart of an existence proof is
obtaining suitable estimates. Then more or less standard approximation
techniques can be used to obtain the desired conclusion. (For a 
discussion of this see \cite{friedrich00a}, section 3.1.) In the problem
studied in \cite{christodoulou00a} it is an appropriate approximation
method which is missing.

One of the problems in tackling the initial value problem for
a dynamical fluid body is that the boundary is moving. It would be very
convenient to use Lagrangian coordinates, since in those coordinates the
boundary is fixed. Unfortunately, it is not at all obvious that the Euler
equations in Lagrangian coordinates have a well-posed initial value problem,
even in the absence of a boundary. It was, however, recently shown by 
Friedrich~\cite{friedrich98b} that it is possible to treat the Cauchy problem 
for fluids in general relativity in Lagrangian coordinates.

In the case of a fluid with non-vanishing boundary density it is not
only the evolution equations which cause problems. It is already difficult
to construct suitable solutions of the constraints. A theorem on this
has recently been obtained by Dain and Nagy \cite{dain02a}. There remains
an undesirable technical restriction, but the theorem nevertheless 
provides a very general class of physically interesting initial data
for a self-gravitating fluid body in general relativity.

\section{Global symmetric solutions}\label{symmetric}

An obvious procedure to obtain special cases of the general global existence
problem for the Einstein equations which are amenable to attack is to make
symmetry assumptions. In this section we discuss the results which have been
obtained for various symmetry classes defined by different choices of number
and character of Killing vectors.

\subsection{Stationary solutions}\label{stationary}

Many of the results on global solutions of the Einstein equations involve
considering classes of spacetimes with Killing vectors. A particularly
simple case is that of a timelike Killing vector, i.~e.\ the case of 
stationary spacetimes. In the vacuum case there are very few solutions
satisfying physically reasonable boundary conditions. This is related
to no hair theorems for black holes and lies outside the scope of this 
review. More information on the topic can be found in the book of 
Heusler~\cite{heusler96} and in his Living Review~\cite{heusler98}.
(See also \cite{beyer01} where the stability of the Kerr metric is discussed.)
The case of phenomenological matter models has been reviewed
in~\cite{rendall97c}. The account given there will be updated in
the following. 

The area of stationary solutions of the Einstein equations coupled to
field theoretic matter models has been active in recent years as a 
consequence of the discovery by Bartnik and McKinnon~\cite{bartnik88b} of a 
discrete family of regular static spherically symmetric solutions of the 
Einstein-Yang-Mills equations with gauge group $SU(2)$. The equations
to be solved are ordinary differential equations and in~\cite{bartnik88b}
they were solved numerically by a shooting method. The first existence proof 
for a solution of this kind is due to Smoller, Wasserman, Yau and McLeod
~\cite{smoller91} and involves an arduous qualitative analysis of the 
differential equations. The work on the Bartnik-McKinnon solutions, including 
the existence theorems, has been extended in many directions. Recently
static solutions of the Einstein-Yang-Mills equations which are not
spherically symmetric were discovered numerically~\cite{kleihaus98}. It is a
challenge to prove the existence of solutions of this kind. Now the
ordinary differential equations of the previously known case are replaced
by elliptic equations. Moreover, the solutions appear to still be discrete, 
so that a simple perturbation argument starting from the spherical case does
not seem feasible. In another development it was shown that a linearized
analysis indicates the existence of stationary non-static solutions
~\cite{brodbeck97}. It would be desirable to study the question of 
linearization stability in this case, which, if the answer were favourable, 
would give an existence proof for solutions of this kind.

Now we return to phenomenological matter models, starting with the case
of spherically symmetric static solutions. Basic existence theorems for
this case have been proved for perfect fluids~\cite{rendall91},
collisionless matter~\cite{rein93},~\cite{rein94a} and elastic bodies
~\cite{park00}. The last of these is the solution to an open problem 
posed in~\cite{rendall97c}. All these theorems demonstrate the existence 
of solutions which are everywhere smooth and exist globally as functions
of area radius for a general class of constitutive relations. The physically 
significant question of the finiteness of the mass of these configurations 
was only answered in these papers under restricted circumstances. For 
instance, in the case of perfect fluids and collisionless matter, solutions 
were constructed by perturbing about the Newtonian case. Solutions for an 
elastic body were obtained by perturbing about the case of isotropic 
pressure, which is equivalent to a fluid. Further progress on the question
of the finiteness of the mass of the solutions was made in the case of a 
fluid by Makino~\cite{makino98}, who gave a rather general criterion on the
equation of state ensuring the finiteness of the radius. Makino's criterion 
was generalized to kinetic theory in~\cite{rein98a}. This resulted in 
existence proofs for various models which have been considered in galactic 
dynamics and which had previously been constructed numerically. 
(Cf.~\cite{binney87},~\cite{shapiro85} for an account of these models in the 
non-relativistic and relativistic cases respectively.) Most of the work
quoted up to now refers to solutions where the support of the density is
a ball. For matter with anisotropic pressure the support may also be a
shell, i.e. the region bounded by two concentric spheres. The existence
of static shells in the case of the Einstein-Vlasov equations was proved
in \cite{rein00a}. 

In the case of self-gravitating Newtonian spherically symmetric 
configurations of collisionless matter, it can be proved that the phase 
space density of particles depends only on the energy of the particle
and the modulus of its angular momentum~\cite{batt86}. This is known as
Jeans' theorem. It was already shown in~\cite{rein94a} that the naive
generalization of this to the general relativistic case does not hold
if a black hole is present. Recently counterexamples to the generalization 
of Jeans' theorem to the relativistic case which are not dependent on a
black hole were constructed by Schaeffer~\cite{schaeffer99}. It remains to 
be seen whether there might be a natural modification of the formulation 
which would lead to a true statement.

For a perfect fluid there are results stating that a static solution is 
necessarily spherically symmetric~\cite{lindblom94}. They still require 
a restriction on the equation of state which it would be desirable to
remove. A similar result is not to be expected in the case of other matter
models, although as yet no examples of non-spherical static solutions are
available. In the Newtonian case examples have been constructed by Rein
~\cite{rein00a}. (In that case static solutions are defined to be those where 
the particle current vanishes.) For a fluid there is an existence theorem
for solutions which are stationary but not static (models for rotating
stars)~\cite{heilig95}. At present there are no corresponding theorems
for collisionless matter or elastic bodies. In~\cite{rein00a} stationary,
non-static configurations of collisionless matter were constructed in the
Newtonian case.

Two obvious characteristics of a spherically symmetric static solution
of the Einstein-Euler equations which has a non-zero density only in
a bounded spatial region are its radius $R$ and its total mass $M$.
For a given equation of state there is a one-parameter family of 
solutions. These trace out a curve in the $(M,R)$ plane. In the 
physics literature there are pictures of this curve which indicate
that it spirals in on a certain point in the limit of large density.
The occurrence of such a spiral and its precise asymptotic form
have been proved rigorously by Makino \cite{makino00a}.  

For some remarks on the question of stability see section~\ref{hydro}.

\subsection{Spatially homogeneous solutions}\label{homogeneous}

A solution of the Einstein equations is called spatially homogeneous if
there exists a group of symmetries with three-dimensional spacelike orbits.
In this case there are at least three linearly independent spacelike
Killing vector fields. For most matter models the field equations reduce
to ordinary differential equations. (Kinetic matter leads to an
integro-differential equation.) The most important results in this area
have been reviewed in a recent book edited by Wainwright and Ellis~\cite{
wainwright97}. See, in particular, part two of the book. There remain a host 
of interesting and accessible open questions. The spatially homogeneous 
solutions have the advantage that it is not necessary to stop at just 
existence theorems; information on the global qualitative behaviour of 
solutions can also be obtained. 

An important question which has been open for a long time concerns the 
mixmaster model, as discussed in~\cite{rendall97d}. This is a class
of spatially homogeneous solutions of the vacuum Einstein equations
which are invariant under the group $SU(2)$. A special subclass of these
$SU(2)$-invariant solutions, the (parameter-dependent) Taub-NUT solution, 
is known explicitly in terms of elementary functions. The Taub-NUT solution 
has a simple initial singularity which is in fact a Cauchy horizon. All 
other vacuum solutions admitting a transitive action of $SU(2)$ on 
spacelike hypersurfaces (Bianchi type IX solutions) will be called
generic in the present discussion. These generic Bianchi IX solutions
(which might be said to constitute the mixmaster solution proper) have 
been believed for a long time to have singularities which are oscillatory
in nature where some curvature invariant blows up. This belief was based
on a combination of heuristic considerations and numerical calculations.
Although these together do make a persuasive case for the accepted picture,
until very recently there were no mathematical proofs of the these 
features of the mixmaster model available. This has now changed. 
First, a proof of curvature blow-up and oscillatory behaviour for a
simpler model (a solution of the Einstein-Maxwell equations) which shares 
many qualitative features with the mixmaster model was obtained by 
Weaver~\cite{weaver99a}. In the much more difficult case of the mixmaster 
model itself corresponding results were obtained by 
Ringstr\"om~\cite{ringstrom00a}. Later he extended this in several
directions in \cite{ringstrom00b}. In that paper more detailed information
was obtained concerning the asymptotics and an attractor for the evolution
was identified. It was shown that generic solutions of Bianchi type IX
with a perfect fluid whose equation of state is $p=(\gamma-1)\rho$
with $1\le\gamma <2$ are approximated near the singularity by vacuum 
solutions. The case of a stiff fluid ($\gamma=2$) which has a
different asymptotic behaviour was analysed completely for all models of
Bianchi class A, a class which includes Bianchi type IX.

Ringstr\"om's analysis of the mixmaster model is potentially of great 
significance for the mathematical understanding of singularities of the
Einstein equations in general. Thus its significance goes far beyond the
spatially homogeneous case. According to extensive investigations of
Belinskii, Khalatnikov and Lifshitz (see~\cite{lifshitz63}, 
~\cite{belinskii70},~\cite{belinskii82} and
references therein) the mixmaster model should provide an approximate 
description for the general behaviour of solutions of the Einstein equations
near singularities. This should apply to many matter models as well as to
the vacuum equations. The work of Belinskii, Khalatnikov and Lifshitz
(BKL) is hard to understand and it is particularly difficult to find a
precise mathematical formulation of their conclusions. This has caused
many people to remain sceptical about the validity of the BKL picture.
Nevertheless, it seems that nothing has ever been found which 
indicates any significant flaws in the final version. As long as the
mixmaster model itself was not understood this represented a fundamental
obstacle to progress on understanding the BKL picture mathematically.
The removal of this barrier opens up an avenue to progress on this issue.
The BKL picture is discussed in more detail in section \ref{sing}.

Some recent and qualitatively new results concerning the asymptotic 
behaviour of spatially homogeneous solutions of the Einstein-matter 
equations, both close to the initial singularity and in a phase of 
unlimited expansion, (and with various matter models) can be found in 
~\cite{rendall99a}, \cite{rendall00a}, \cite{rendall01b}, 
\cite{wainwright99a}, \cite{nilsson00a} and \cite{hewitt01}. These 
show in particular that 
the dynamics can depend sensitively on the form of matter chosen. (Note
that these results are consistent with the BKL picture.) The dynamics of
indefinitely expanding cosmological models is discussed further in 
section \ref{expand}. 

\subsection{Spherically symmetric solutions}\label{spherical}

The most extensive results on global inhomogeneous solutions of the
Einstein equations obtained up to now concern spherically symmetric
solutions of the Einstein equations coupled to a massless scalar 
field with asymptotically flat initial data.
In a series of papers Christodoulou~\cite{christodoulou86a,
christodoulou86b,christodoulou87a,christodoulou87b,christodoulou91,
christodoulou93a,christodoulou94,christodoulou99} has proved a
variety of deep results on the global structure of these solutions.
Particularly notable are his proofs that naked singularities can 
develop from regular initial data~\cite{christodoulou94} and that this 
phenomenon is unstable with respect to perturbations of the data 
~\cite{christodoulou99}. In related work Christodoulou 
~\cite{christodoulou95,christodoulou96a,christodoulou96b} has studied 
global spherically symmetric solutions of the Einstein equations coupled to 
a fluid with a special equation of state (the so-called two-phase model).
A generalization of the results of \cite{christodoulou86a} to the case
of a nonlinear scalar field has been given by Chae \cite{chae01a}. 

The rigorous investigation of the spherically symmetric collapse of 
collisionless matter in general relativity was initiated by Rein and the author
~\cite{rein92}, who showed that the evolution of small initial data leads to
geodesically complete spacetimes where the density and curvature fall off 
at large times. Later it was shown~\cite{rein95a} that independent of the 
size of the initial data the first singularity, if there is one at all,
must occur at the centre of symmetry. This result uses a time coordinate of
Schwarzschild type; an analogous result for a maximal time coordinate was
proved in~\cite{rendall97e}. The question of what happens for general large
initial data could not yet be answered by analytical techniques. In 
~\cite{rein98b} numerical methods were applied in order to try to make some 
progress in this direction. The results are discussed in the next paragraph.

Despite the range and diversity of the results obtained by Christodoulou
on the spherical collapse of a scalar field, they do not encompass some
of the most interesting phenomena which have been observed numerically.
These are related to the issue of critical collapse. For sufficiently small 
data the field disperses. For sufficiently large data a black hole is 
formed. The question is what happens in between. This can be investigated 
by examining a one-parameter family of initial data interpolating between
the two cases. It was found by Choptuik~\cite{choptuik93} that there is a
critical value of the parameter below which dispersion takes place and above
which a black hole is formed and that the mass of the black hole approaches
zero as the critical parameter value is approached. This gave rise to a
large literature where the spherical collapse of different kinds of matter
was computed numerically and various qualitative features were determined.
For reviews of this see~\cite{gundlach98} and~\cite{gundlach99}. In the 
calculations of ~\cite{rein98b} for collisionless matter it was found
that in the situations considered the black hole mass tended to a strictly
positive limit as the critical parameter was approached from above. These
results were confirmed and extended by Olabarrieta and Choptuik
\cite{olabarrieta02a}. There are no rigorous mathematical results available 
on the issue of a mass gap for either a scalar field or collisionless matter 
and it is an outstanding challenge for mathematical relativists to change 
this situation. 

Another aspect of Choptuik's results is the occurrence of a discretely
self-similar solution. It would seem hard to prove the existence of a
solution of this kind analytically. For other types of matter, such as
a perfect fluid with linear equation of state, the critical
solution is continuously self-similar and this looks more tractable. The
problem reduces to solving a system of singular ordinary differential 
equations subject to certain boundary conditions. This problem 
was solved in~\cite{christodoulou94} for the case where the matter
model is given by a massless scalar field, but the solutions produced there,
which are continuously self-similar, cannot include the Choptuik critical 
solution. Bizo\'n and Wasserman \cite{bizon02a} studied the corresponding 
problem for the Einstein equations coupled to a wave map with target 
$SU(2)$. They proved the existence of continuously self-similar solutions
including one which, according the results of numerical calculations, 
appears to play the role of critical solution in collapse. Another
case where the question of the existence of the critical solution seems 
to be a problem which could possibly be solved in the near 
future is that of a perfect fluid. A good starting point for this is the 
work of Goliath, Nilsson and Uggla~\cite{goliath98a},~\cite{goliath98b}. 
These authors gave a formulation of the problem in terms of dynamical 
systems and were able to determine certain qualitative features of the 
solutions. See also \cite{carr00a} and \cite{carr01a}.

A possible strategy for learning more about critical collapse, pursued by
Bizo\'n and collaborators, is to study model problems in flat space which 
exhibit features similar to those observed numerically in the case of the 
Einstein equations. Up to now only models showing continuous 
self-similarity have been found. These include wave maps in various 
dimensions and the Yang-Mills equations in spacetimes of dimension greater 
than four. As mentioned in section \ref{differentiability} it is known that 
in four dimensions there exist global smooth solutions of the Yang-Mills
equations corresponding to rather general initial data
\cite{eardley82}, \cite{klainerman95}. In dimensions greater than five
it is known that there exist solutions which develop singularities in
finite time. This follows from the existence of continuously self-similar
solutions \cite{bizon01b}. Numerical evidence indicates that this type 
of blow-up is stable, i.e. occurs for an open set of initial data. The
numerical work also indicates that there is a critical self-similar 
solution separating this kind of blow-up from dispersion. The spacetime
dimension five is critical for Yang-Mills theory. Apparently singularities
form, but in a different way from what happens in dimension six. There
is as yet no rigorous proof of blow-up in five dimensions. 

The effects 
found in Yang-Mills theory are mirrored in two dimensions less by wave maps
with values in spheres \cite{bizon01a}. In four dimensions blow-up is known 
while in three dimensions there appears (numerically) to be a kind of blow-up 
similar to that found for Yang-Mills in dimension five. There is no 
rigorous proof of blow-up. What is seen numerically is that the collapse
takes place by scaling within a one-parameter family of static solutions.
The case of wave maps is the most favourable known model problem for 
proving theorems about critical phenomena associated to singularity
formation. The existence of a solution having the properties expected
of the critical solution for wave maps in four dimensions has been proved 
in \cite{bizon00a}. Some rigorous support for the numerical findings in 
three dimensions has been given by work of Struwe. (See the preprints 
available from \cite{struwepreprints}.) He showed among other things
that if there is blow-up in finite time it must take place in a way
resembling that observed in the numerical calculations.  

Self-similar solutions are characteristic of what is called Type II critical
collapse. In Type I collapse an analogous role is played by static solutions
and quite a bit is known the existence of these. For instance in the case
of the Einstein-Yang-Mills equations it is one of the Bartnik-McKinnon
solutions mentioned in section \ref{stationary} which does this. In 
the case of collisionless matter the results of \cite{olabarrieta02a}
show that at least in some cases critical collapse is mediated by a static
solution in the form of a shell. There are existence results for shells
of this kind \cite{rein99a} although no connection has yet been made 
between those shells whose existence been proved and those which have
been observed numerically in critical collapse calculations. Note that
Mart\'\i n-Garc\'\i a and Gundlach\cite{martingarcia01} have presented a 
(partially numerical) construction of self-similar solutions of the 
Einstein-Vlasov system. 

\subsection{Cylindrically symmetric solutions}\label{cylindrical} 

Solutions of the Einstein equations with cylindrical symmetry which are
asymptotically flat in all directions allowed by the symmetry represent an
interesting variation on asymptotic flatness. Since black holes are
apparently incompatible with this symmetry, one may hope to prove geodesic 
completeness of solutions under appropriate assumptions. (It would be
interesting to have a theorem making the statement about black holes
precise.) A proof of geodesic completeness has been achieved
for the Einstein vacuum equations and for the source-free Einstein-Maxwell
equations in~\cite{berger95}, building on global existence theorems for
wave maps~\cite{zadeh93a,zadeh93b}. For a quite different point of view 
on this question involving integrable systems see~\cite{woodhouse97}.
A recent paper of Hauser and Ernst~\cite{hauser01} also appears to
be related to this question. However, due to the great length of this text
and its reliance on many concepts unfamiliar to this author, no further
useful comments on the subject can be made here.    

\subsection{Spatially compact solutions}\label{compact}

In the context of spatially compact spacetimes it is first necessary
to ask what kind of global statements are to be expected. In a
situation where the model expands indefinitely it is natural to
pose the question whether the spacetime is causally geodesically complete
towards the future. In a situation where the model develops a singularity
either in the past or in the future one can ask what the qualitative
nature of the singularity is. It is very difficult to prove results of
this kind. As a first step one may prove a global existence theorem in
a well-chosen time coordinate. In other words, a time coordinate is chosen
which is geometrically defined and which, under ideal circumstances, will
take all values in a certain interval $(t_-,t_+)$. The aim is then to show 
that, in the maximal Cauchy development of data belonging to a certain class, 
a time coordinate of the given type exists and exhausts the expected 
interval. The first result of this kind for inhomogeneous spacetimes was
proved by Moncrief in~\cite{moncrief81b}. This result concerned Gowdy 
spacetimes. These are vacuum spacetimes with a two-dimensional Abelian 
group of isometries  
acting on compact orbits. The area of the orbits defines a natural time 
coordinate (areal time coordinate). Moncrief showed that in the maximal 
Cauchy development of data given on a hypersurface of constant time, this 
time coordinate takes on the maximal possible range, namely $(0,\infty).$ 
This result was extended to more general vacuum spacetimes with two Killing 
vectors in~\cite{berger97}. Andr\'easson~\cite{andreasson99} extended it in 
another direction to the case of collisionless matter in a spacetime with 
Gowdy symmetry.

Another attractive time coordinate is constant mean curvature (CMC) time.
For a general discussion of this see~\cite{rendall96a}. A global existence 
theorem in this time for spacetimes with two Killing vectors and certain 
matter models (collisionless matter, wave maps) was proved in 
~\cite{rendall97b}. That the choice of matter model is important for this 
result was demonstrated by a global non-existence result for dust in 
~\cite{rendall97a}. As shown in~\cite{isenberg98}, this leads to 
examples of spacetimes which are not covered by a CMC slicing.  
Results on global existence of CMC foliations have also been obtained 
for spherical and hyperbolic symmetry 
~\cite{rendall95a, burnett96}. 

A drawback of the results on the existence of CMC foliations just cited is
that they require as a hypothesis the existence of one CMC Cauchy surface
in the given spacetime. More recently this restriction has been removed 
in certain cases by Henkel using a generalization of CMC foliations called
prescribed mean curvature (PMC) foliations. A PMC foliation can be built 
which includes any given
Cauchy surface \cite{henkel01a} and global existence of PMC foliations
can be proved in a way analogous to that previously done for CMC 
foliations \cite{henkel01b, henkel01c}. These global foliations provide 
barriers which imply the existence of a CMC hypersurface. Thus in the end 
it turns out that the unwanted condition in the previous theorems on CMC 
foliations is in fact automatically satisfied. Connections between  
areal, CMC and PMC time coordinates were further explored in
\cite{andreasson01a}. One important observation there is that
hypersurfaces of constant areal time in spacetimes with symmetry often
have mean curvature of a definite sign. 

Once global existence has been proved for a preferred time coordinate, the
next step is to investigate the asymptotic behaviour of the solution as 
$t\to t_{\pm}$. There are few cases in which this has been done successfully.
Notable examples are Gowdy spacetimes~\cite{chrusciel90a, isenberg90,
chrusciel90b} and solutions of the Einstein-Vlasov system with spherical
and plane symmetry~\cite{rein96a}. Progress in constructing spacetimes with
prescribed singularities will be described in section~\ref{prescribe}. In the 
future 
this could lead in some cases to the determination of the asymptotic behaviour
of large classes of spacetimes as the singularity is approached.

\section{Newtonian theory and special relativity}\label{newtonian}

To put the global results discussed in this article into context it is helpful
to compare with Newtonian theory and special relativity. Some of the 
theorems which have been proved in those contexts and which can offer
insight into questions in general relativity will now be reviewed. It
should be noted that even in these simpler contexts open questions abound.

\subsection{Hydrodynamics}\label{hydro} 

Solutions of the classical (compressible) Euler equations typically 
develop singularities, i.~e.\ discontinuities of the basic fluid variables,
in finite time~\cite{sideris79}. Some of the results of~\cite{sideris79}
were recently generalized to the case of a relativistic 
fluid~\cite{guo99a}. The proofs of the development of singularities are
by contradiction and so do not give information about what happens when
the smooth solution breaks down. One of the things which can happen is
the formation of shock waves and it is known that at least in certain
cases solutions can be extended in a physically meaningful way beyond the 
time of shock formation. The
extended solutions only satisfy the equations in the weak sense. For 
the classical Euler equations there is a well-known theorem on global
existence of weak solutions in one space dimension which goes back
to~\cite{glimm65}. This has been generalized to the relativistic case.
Smoller and Temple treated the case of an isentropic fluid with linear
equation of state~\cite{smoller93} while Chen analysed the cases of
polytropic equations of state~\cite{chen95} and flows with variable entropy
~\cite{chen97}. This means that there is now an understanding of this
question in the relativistic case similar to that available in the
classical case. 

In space dimensions higher than one there are no general global existence
theorems. For a long time there were also no uniqueness theorems for
weak solutions even in one dimension. It should be emphasized that weak
solutions can easily be shown to be non-unique unless they are required
to satisfy additional restrictions such as entropy conditions. A reasonable
aim is to find a class of weak solutions in which both existence and 
uniqueness hold. In the one-dimensional case this has recently been achieved 
by Bressan and collaborators (see~\cite{bressan95a},~\cite{bressan95b},
\cite{bressan00a} and references therein).

It would be desirable to know more about which quantities must blow up
when a singularity forms in higher dimensions. A partial answer was 
obtained for classical hydrodynamics by Chemin~\cite{chemin90}. The
possibility of generalizing this to relativistic and self-gravitating
fluids was studied by Brauer~\cite{brauer95}. There is one situation in which 
a smooth solution of the classical Euler equations is known to exist for all
time. This is when the initial data are small and the fluid initially 
flowing uniformly outwards. A theorem of this type has been proved by
Grassin~\cite{grassin98}. There is also a global existence result due to
Guo~\cite{guo98a} for an irrotational charged fluid in Newtonian physics, 
where the repulsive effect of the charge can suppress the formation of 
singularities.

A question of great practical interest for physics is that of the stability
of equilibrium stellar models. Since, as has already been pointed out, we
know so little about the global time evolution for a self-gravitating fluid
ball, even in the Newtonian case, it is not possible to say anything 
rigorous about nonlinear stability at the present time. We can, however,
make some statements about linear stability. The linear stability of a large
class of static spherically symmetric solutions of the Einstein-Euler
equations within the class of spherically symmetric perturbations has been 
proved by Makino~\cite{makino98}. (Cf.\ also~\cite{lin97} for the Newtonian
problem.) The spectral properties of the linearized
operator for general (i.~e.\ non-spherically symmetric) perturbations in the
Newtonian problem have been studied by Beyer~\cite{beyer95}.
This could perhaps provide a basis for a stability analysis, but this
has not been done. 

\subsection{Kinetic theory}

Collisionless matter is known to admit a global singularity-free
evolution in many cases. For self-gravitating collisionless matter,
which is described by the Vlasov-Poisson system, there is a general 
global existence theorem~\cite{pfaffelmoser92},~\cite{lions91}. There
is also a version of this which applies to Newtonian cosmology~\cite{rein94b}.
A more difficult case is that of the Vlasov-Maxwell system, which
describes charged collisionless matter. Global existence is not known
for general data in three space dimensions but has been shown
in two space dimensions~\cite{glassey98a},~\cite{glassey98b} and in three
dimensions with one symmetry~\cite{glassey97} or with almost spherically 
symmetric data~\cite{rein90}.

The nonlinear stability of static solutions of the Vlasov-Poisson system 
describing Newtonian self-gravitating collisionless matter has been 
investigated using the energy-Casimir method. For information on this 
see~\cite{guo01a} and its references. The energy-Casimir method has
been applied to the Einstein equations in \cite{wolansky01a}.

For the classical Boltzmann equation global existence and uniqueness
of smooth solutions has been proved for homogeneous initial data and 
for data which are small or close to equilibrium. For general data with 
finite energy and entropy global existence of weak solutions (without 
uniqueness) was proved by DiPerna and Lions~\cite{diperna89}. For 
information on these results and on the classical Boltzmann equation in 
general see~\cite{cercignani88},~\cite{cercignani94}. Despite the 
non-uniqueness it is possible to show that all solutions tend to 
equilibrium at late times. This was first proved by Arkeryd~\cite{arkeryd92} 
by non-standard analysis and then by Lions~\cite{lions94} without those 
techniques. It should be noted that since the usual conservation laws for
classical solutions are not known to hold for the DiPerna-Lions solutions,
it is not possible to predict which equilibrium solution a given solution
will converge to. In the meantime analogues of several of these results for
the classical Boltzmann equation have been proved in the relativistic case. 
Global existence of weak solutions was proved in~\cite{dudynski92}. Global 
existence and convergence to equilibrium for classical solutions starting 
close to equilibrium was proved in~\cite{glassey93}. On the other hand global
existence of classical solutions for small initial data is not known.
Convergence to equilibrium for weak solutions with general data was proved
by Andr\'easson~\cite{andreasson96}. There is still no existence and uniqueness
theorem in the literature for general spatially homogeneous solutions of the 
relativistic Boltzmann equation. (A paper claiming to prove existence and
uniqueness for solutions of the Einstein-Boltzmann system which are 
homogeneous and isotropic~\cite{mucha99} contains fundamental errors.)

\subsection{Elasticity theory}

There is an extensive literature on mathematical elasticity theory but
the mathematics of self-gravitating elastic bodies seems to have been 
largely neglected. An existence theorem for spherically symmetric 
elastic bodies in general relativity was mentioned in section 
\ref{stationary}. More recently Beig and Schmidt \cite{beig02a} proved an 
existence theorem for static elastic bodies subject to Newtonian gravity 
which need not be spherically symetric.

\section{Global existence for small data}\label{small}

An alternative to symmetry assumptions is provided by \lq small data\rq\ 
results, where solutions are studied which develop from data close to 
those for known solutions. This leads to some simplification in comparison
to the general problem, but with present techniques it is still very hard
to obtain results of this kind.

\subsection{Stability of de Sitter space}\label{desitter}

In~\cite{friedrich86} Friedrich proved a result on the stability of de Sitter 
space. He gives data at infinity but the same type of argument can be applied
starting from a Cauchy surface in spacetime to give an analogous result.
This concerns the Einstein vacuum equations with positive cosmological
constant and is as follows. Consider initial data induced by
de Sitter space on a regular Cauchy hypersurface. Then all initial
data (vacuum with positive cosmological constant) near enough to
these data in a suitable (Sobolev) topology have maximal Cauchy
developments which are geodesically complete. The result gives
much more detail on the asymptotic behaviour than just this and may
be thought of as proving a form of the cosmic no hair conjecture in the
vacuum case. (This conjecture says roughly that the de Sitter solution
is an attractor for expanding cosmological models with positive
cosmological constant.) This result is proved using conformal techniques
and, in particular, the regular conformal field equations developed by
Friedrich.

There are results obtained using the regular conformal field equations
for negative or vanishing cosmological constant~\cite{friedrich95,
friedrich98a} but a detailed discussion of their nature would be out of place 
here. (Cf.\ however section~\ref{hyperboloidal}.)

\subsection{Stability of Minkowski space}\label{minkowski}

Another result on global existence for small data is that of Christodoulou
and Klainerman on the stability of Minkowski space \cite{christodoulou93b} 
The formulation of the result is close to that given in section
\ref{desitter} but now de Sitter space is replaced by Minkowski space. 
Suppose then that initial data for the vacuum Einstein equations
are prescribed which are asymptotically flat and sufficiently close to
those induced by Minkowski space on a hyperplane. Then Christodoulou and
Klainerman prove that the maximal Cauchy development of these data is
geodesically complete. They also provide a wealth of detail on the
asymptotic behaviour of the solutions. The proof is very long and technical.
The central tool is the Bel-Robinson tensor which plays an analogous role 
for the gravitational field to that played by the energy-momentum tensor 
for matter fields. Apart from the book of Christodoulou and Klainerman
itself some introductory material on geometric and analytic aspects of the
proof can be found in~\cite{bourguignon92} and~\cite{christodoulou90} 
respectively. More recently the result for the vacuum Einstein equations 
has been generalized to the case of the Einstein-Maxwell system by Zipser
\cite{zipserpreprint}.

In the original version of the theorem initial data had to be prescribed on
all of ${\bf R}^3$. A generalization described in~\cite{klainerman99} concerns
the case where data need only be prescribed on the complement of a compact 
set in ${\bf R}^3$. This means that statements can be obtained for any 
asymptotically flat spacetime where the initial matter distribution has 
compact support, provided attention is confined to a
suitable neighbourhood of infinity. The proof of the new version uses
a double null foliation instead of the foliation by spacelike hypersurfaces
previously used and leads to certain conceptual simplifications.

\subsection{Stability of the (compactified) Milne model}\label{milne} 

The interior of the light cone in Minkowski space foliated by the 
spacelike hypersurfaces of constant Lorentzian distance from the origin
can be thought of as a vacuum cosmological model, sometimes known as
the Milne model. By means of a suitable discrete subgroup of the Lorentz
group it can be compactified to give a spatially compact cosmological
model. With a slight abuse of terminology the latter spacetime will also 
be referred to here as the Milne model. A proof of the stability of the
latter model by Andersson and Moncrief has been announced in 
~\cite{andersson99a}. The result is that, given data for the Milne model
on a manifold obtained by compactifying a hyperboloid in Minkowski space,
the maximal Cauchy developments of nearby data are geodesically complete
in the future. Moreover the Milne model is asymptotically stable in
the sense that any other solution in this class converges towards the
Milne model in terms of suitable dimensionless variables.

The techniques used by Andersson and Moncrief are similar to those used
by Christodoulou and Klainerman. In particular, the Bel-Robinson tensor
is crucial. However their situation is much simpler than that of 
Christodoulou and Klainerman, so that the complexity of the proof is
not so great. This has to do with the fact that the fall-off of the 
fields towards infinity in the Minkowksi case  is different in different 
directions, while it is uniform in the Milne case. Thus it is enough in
the latter case to always contract the Bel-Robinson tensor with the
same timelike vector when deriving energy estimates. The fact that the 
proof is simpler opens up a real possibility of generalizations, for
instance by adding different matter models.   

\subsection{Stability of the Bianchi type III form of flat 
spacetime}\label{bianchi3}

Another vacuum cosmological model whose nonlinear stability has been
investigated is the Bianchi III form of flat spacetime. To get this model
first do the construction described in the last section with the difference
that the starting solution is three-dimensional Minkowski space. Then
take the metric product of the resulting three-dimensional Lorentz
manifold with a circle. This defines a flat spacetime which has one
Killing vector which is the generator of rotations of the circle.
It has been shown by Choquet-Bruhat and Moncrief \cite{choquet01a} that this 
solution is stable under small vacuum perturbations preserving the one 
dimensional symmetry. More precisely, the result is proved only for the
polarized case, but the authors suggest that this restriction can be lifted
at the expense of doing some more work. As in the case of the Milne model, 
a natural task is to generalize this result to spacetimes with suitable 
matter content. The reasons why it is necessary to restrict to symmetric 
perturbations in this analysis, in contrast to what happens with the Milne 
model, is discussed in detail in \cite{choquet01a}.

One of the main techniques used is a method of modified energy estimates
which is likely to be of more general applicability. The Bel-Robinson
tensor plays no role. The other main technique is based on the fact that
the problem under study is equivalent to the study of the 2+1-dimensional
Einstein equations coupled to a wave map (a scalar field in the polarized
case). This helps to explain why the use of the Dirichlet energy could be
imported into this problem from the work of \cite{andersson97a} on 
2+1 vacuum gravity.  

\section{Prescribed singularities}\label{prescribe}

If it is too hard to get information on the qualitative nature of solutions
by evolving from a regular initial hypersurface towards a possible 
singularity an alternative approach is to construct spacetimes with given
singularities. Recently the latter method has made significant progress
and the new results are presented in this section.

\subsection{Isotropic singularities}\label{isotropic}

The existence and uniqueness results discussed in this section are
motivated by Penrose's Weyl curvature hypothesis. Penrose suggests
that the initial singularity in a cosmological model should be such
that the Weyl tensor tends to zero or at least remains bounded. There
is some difficulty in capturing this by a geometric condition and it
was suggested in~\cite{goode85} that a clearly formulated geometric condition
which, on an intuitive level, is closely related to the original
condition, is that the conformal structure should remain regular at the 
singularity. Singularities of this type are known as conformal or
isotropic singularities.

Consider now the Einstein equations coupled to a perfect fluid with the
radiation equation of state $p=\rho/3$. Then it has been shown~\cite{newman93,
claudel98} that solutions with an isotropic singularity are determined
uniquely by certain free data given at the singularity. The data which 
can be given is, roughly speaking, half as much as in the case of a 
regular Cauchy hypersurface. The method of proof is to derive an existence
and uniqueness theorem for a suitable class of singular hyperbolic equations.
In~\cite{anguige99a} this was extended to the equation of state 
$p=(\gamma-1)\rho$ for any $\gamma$ satisfying $1<\gamma\le 2$.

What happens to this theory when the fluid is replaced by a different
matter model? The study of the case of a collisionless gas of massless
particles was initiated
in~\cite{anguige99b}. The equations were put into a form similar to that 
which was so useful in the fluid case and therefore likely to be
conducive to proving existence theorems. Then theorems of this kind were
proved in the homogeneous special case. These were extended to the general
(i.~e.\ inhomogeneous) case in~\cite{anguige00b}. The picture obtained for
collisionless matter is very different from that for a perfect fluid. Much
more data can be given freely at the singularity in the collisionless case.

These results mean that the problem of isotropic singularities has largely
been solved. There do, however, remain a couple of open questions. What
happens if the massless particles are replaced by massive ones? What happens
if the matter is described by the Boltzmann equation with non-trivial
collision term? Does the result in that case look more like the Vlasov
case or more like the Euler case?

\subsection{Fuchsian equations}\label{fuchsian}

The singular equations which arise in the study of isotropic singularities
are closely related to what Kichenassamy~\cite{kichenassamy96a} calls
Fuchsian equations. He has developed a rather general theory of these
equations. (See~\cite{kichenassamy96a},~\cite{kichenassamy96b}, 
~\cite{kichenassamy96c}, and also the earlier papers~\cite{baouendi77},
~\cite{kichenassamy93a} and~\cite{kichenassamy93b}.) In 
~\cite{kichenassamy98a} this was applied to analytic Gowdy spacetimes 
on $T^3$ to construct a family of vacuum spacetimes depending on
the maximum number of free functions (for the given symmetry class) whose
singularities can be described in detail. The symmetry assumed in that paper
requires the two-surfaces orthogonal to the group orbits to be 
surface-forming (vanishing twist constants). In~\cite{isenberg99a} a 
corresponding result was obtained for the class of vacuum spacetimes with 
polarized  $U(1)\times U(1)$ symmetry and non-vanishing twist. The
analyticity requirement on the free functions in the case of Gowdy 
spacetimes on $T^3$ was reduced to smoothness in \cite{rendall00b}. 
There are also Gowdy spacetimes on $S^3$ and $S^2\times S^1$ which have
been less studied than those on $T^3$. The Killing vectors have zeros,
defining axes, and these lead to technical difficulties.
In \cite{stahl01a} Fuchsian techniques were applied to Gowdy spacetimes
on $S^3$ and $S^2\times S^1$. The maximum number of free functions was not 
obtained due to difficulties on the axes. 

Anguige~\cite{anguige00a} has obtained results on solutions with perfect 
fluid which are general under the condition of plane symmetry, which is
stronger than Gowdy symmetry. He also extended this to polarized Gowdy
symmetry in \cite{anguige00c}.

Work related to these Fuchsian methods was done earlier in a somewhat 
simpler context by Moncrief~\cite{moncrief82} who showed the existence of 
a large class of analytic vacuum spacetimes with Cauchy horizons.

As a result of the BKL picture it cannot be expected that the singularities
in general solutions of the Einstein equations in vacuum or with a non-stiff 
fluid can be handled using Fuchsian techniques. (Cf. section \ref{homsing}.)
However things look better in the presence of a massless scalar field or
a stiff fluid. For these types of matter it has been possible 
\cite{andersson01a} to prove a theorem analogous to that of 
\cite{kichenassamy98a} without requiring symmetry assumptions. 
The same conclusion can be obtained for a scalar field with mass or with
a potential of moderate growth \cite{rendall00c}.

The results included in this review concern the Einstein equations in 
four spacetime dimensions. Of course many of the questions discussed have
analogues in other dimensions and these may be of interest for string
theory and related topics. In \cite{damour02a} Fuchsian techniques were
applied to the Einstein equations coupled to a variety of field theoretic
matter models in arbitrary dimensions. One of the highlights is the result
that it is possible to apply Fuchsian techniques without requiring symmetry
assumptions to the vacuum Einstein equations in spacetime dimension at 
least eleven. Many new results are also obtained in four dimensions.
For instance the Einstein-Maxwell-dilaton and Einstein-Yang-Mills
equations are treated. The general nature of the results is that 
provided certain inequalities are satisfied by coupling constants
solutions with prescribed singularities can be constructed which 
depend on the same number of free functions as the general solution
of the given Einstein-matter system.  

\section{Asymptotics of expanding cosmological models}\label{expand}

The aim of this section is to present a picture of the dynamics of
forever expanding cosmological models, by which we mean spacetimes
which are maximal globally hyperbolic developments and which can
be covered by a foliation by Cauchy surfaces whose mean curvature
$\tr k$ is strictly negative. In contrast to the approach to the big 
bang considered in section \ref{sing} the spatial topology can be
expected to play an important role in the present considerations.
Intuitively, it may well happen that gravitational waves have time
to propagate all the way around the universe. It will be assumed,
as the simplest case, that the spacetimes considered admit a compact 
Cauchy surface. Then the hypersurfaces of negative mean curvature
introduced above have finite volume and this volume is a strictly
increasing function of time.

\subsection{Lessons from homogeneous solutions}

Which features should we focus on when thinking about the dynamics
of forever expanding cosmological models? Consider for a moment the
Kasner solution
\begin{equation}\label{kasner}
-dt^2+t^{2p_1}dx^2+t^{2p_2}dy^2+t^{2p_3}dz^2
\end{equation} 
where $p_1+p_2+p_3=1$ and $p_1^2+p_2^2+p_3^2=1$. These are the first 
and second Kasner relations. They imply that not all $p_i$ can be
strictly positive. Taking the coordinates $x$, $y$ and $z$ to be
periodic gives a vacuum cosmological model whose spatial topology is that
of a three-torus. The volume of the hypersurfaces $t=$const. grows
monotonically. However the geometry does not expand in all
directions, since not all $p_i$ are positive. This can be reformulated
in a way which is more helpful when generalizing to inhomogeneous
models. In fact the quantities $-p_i$ are the eigenvalues of the
second fundamental form. The statement then is that the second
fundamental form is not negative definite. Looking at other
homogeneous models indicates that this behaviour of the Kasner
solution is not typical of what happens more generally. On the
contrary, it seems reasonable to conjecture that in general
the second fundamental form eventually becomes negative definite,
at least in the presence of matter.

Some examples will now be presented. The following discussion makes use
of the Bianchi classification of homogenous cosmological models (see e.g.
\cite{wainwright97}). If we take the Kasner solution
and add a perfect fluid with equation of state $p=(\gamma-1)\rho$, 
$1\le\gamma <2$, maintaining the symmetry (Bianchi type I)
then the eigenvalues $\lambda_i$ of the second fundamental
satisfy $\lambda_i/\tr k\to 1/3$ in the limit of infinite expansion.
The solution isotropizes. More generally this does not happen. If
we look at models of Bianchi type II with non-tilted perfect fluid,
i.e. where the fluid velocity is orthogonal to the homogeneous 
hypersurfaces, then the quantities $p_i=\lambda_i/\tr k$ converge
to limits which are positive but differ from $1/3$ (see \cite{wainwright97},
p. 138.) There is 
partial but not complete isotropization. The quantities $p_i$
just introduced are called generalized Kasner exponents, since
in the case of the Kasner solution they reduce to the $p_i$ in
the metric form (\ref{kasner}). This kind of partial isotropization,
ensuring the definiteness of the second fundamental form at late
times, seems to be typical.

Intuitively, a sufficiently general vacuum spacetime should
resemble gravitational waves propagating on some metric describing
the large-scale geometry. This could even apply to spatially
homogeneous solutions, provided they are sufficiently general.
Hence in that case also there should be partial isotropization.
This expectation is confirmed in the case of vacuum spacetimes of 
Bianchi type VIII \cite{ringstrom01a}. In that case the generalized
Kasner exponents converge to non-negative limits different from $1/3$. 
For a vacuum model this can only happen if the quantity
$\hat R=R/(\tr k)^2$, where $R$ is the spatial scalar curvature,
does not tend to zero in the limit of large time.

The Bianchi models of type VIII are the most general indefinitely 
expanding models of class A. Note, however, that models of class VI${}_h$
for all $h$ together are just as general. The latter models with perfect
fluid and equation of state $p=(\gamma-1)\rho$ sometimes tend to the 
Collins model for an open set of values of $h$ for each fixed $\gamma$ (cf. 
\cite{wainwright97}, p. 160). These models do not in general exhibit partial 
isotropization. It is interesting to ask whether this is connected to the 
issue of spatial boundary conditions. General models of class B cannot be 
spatially compactified in such a way as to be locally spatially homogeneous 
while models of Bianchi type VIII can. See also the discussion in 
\cite{barrow01a}.

Another issue is what assumptions on matter are required in order
that it have the effect of (partial) isotropization. Consider the
case of Bianchi I. The case of a perfect fluid has already been
mentioned. Collisionless matter described by kinetic theory also leads 
to isotropization (at least under the assumption of reflection symmetry),
as do fluids with almost any physically reasonable
equation of state \cite{rendall96b}. There is, however, one exception.
This is the stiff fluid, which has a linear equation of state with
$\gamma=2$. In that case the generalized Kasner exponents are 
time-independent, and may take on negative values. In a model with
two non-interacting fluids with linear equation of state the one
with the smaller value of $\gamma$ dominates the dynamics at late
times \cite{coley92a} and so the isotropization is restored. Consider
now the case of a magnetic field and a perfect fluid with linear equation
of state. A variety of cases of Bianchi types I, II and VI${}_0$ have been 
studied in \cite{leblanc97a}, \cite{leblanc98a} and \cite{leblanc95a}, 
with a mixture
of rigorous results and conjectures being obtained. The general picture
seems to be that, apart from very special cases, there is at least partial
isotropization. The asymptotic behaviour varies with the parameter $\gamma$
in the equation of state and with the Bianchi type. (Only the case 
$\gamma\ge 1$ will be considered here.) At one extreme, Bianchi
type I models with $\gamma\le 4/3$ isotropize. At the other extreme the
long time behaviour resembles that of a magnetovacuum model. This occurs
for $\gamma>5/3$ in type I, for $\gamma>10/7$ in type II and for 
all $\gamma>1$ in type VI${}_0$. In all these cases there is partial
isotropization.

Under what circumstances can a spatially homogeneous spacetime
have the property that the generalized Kasner exponents are 
independent of time? The strong energy condition says that
$R_{\alpha\beta}n^\alpha n^\beta\ge 0$ for any causal vector
$n^\alpha$. It follows from the Hamiltonian constraint and the 
evolution equation for $\tr k$ that if the generalized Kasner
exponents are constant in time in a spacetime of Bianchi type I 
then the normal vector $n^\alpha$ to the homogeneous hypersurfaces 
gives equality in the inequality of the strong energy condition.
Hence the matter model is in a sense on the verge of violating
the strong energy condition and this is a major restriction
on the matter model. 

A further question which can be posed concerning the dynamics of
expanding cosmological models is whether $\hat\rho=\rho/(\tr k)^2$
tends to zero. This is of cosmological interest since $\hat\rho$
is (up to a constant factor) the density parameter $\Omega$ used
in the cosmology literature. Note that it is not hard to show that
$\tr k$ and $\rho$ each tend to zero in the limit for any model
with $\Lambda=0$ which exists globally in the future and where the 
matter satisfies the dominant and strong energy conditions. First
it can be seen from the evolution equation for $\tr k$ that this
quantity is monotone increasing and tends to zero as $t\to\infty$.
Then it follows from the Hamiltonian constraint that $\rho$ tends
to zero.

A reasonable condition to be demanded of an expanding cosmological model
is that it be future geodesically complete. This has been proved for many
homogeneous models in \cite{rendall95c}

\subsection{Inhomogeneous solutions with $\Lambda=0$}

For inhomogeneous models with vanishing cosmological constant
there is little information available about
what happens in general. Fischer and Moncrief \cite{fischer99a} have made
a interesting proposal which attempts to establish connections
between the evolution of a suitably conformally rescaled version of 
the spatial metric in an expanding cosmological model and themes in
Riemannian geometry such as the Thurston geometrization conjecture
\cite{thurston97a}, degeneration of families of metrics with bounded curvature
\cite{anderson97a} and the Ricci flow \cite{hamilton82a}. A key element of 
this picture is the theorem on the stability of the Milne model discussed in
section \ref{milne}. More generally the rescaled metric is supposed
to converge to a hyperbolic metric (metric of constant negative curvature) 
on a region which is large in the sense that the volume of
its complement tends to zero. If the topology of the Cauchy surface
is such that it is consistent with a metric of some Bianchi type
then the hyperbolic region will be missing and the volume of the
entire rescaled metric will tend to zero. In this situation it
might be expected that the metric converges to a (locally) 
homogeneous metric in some sense. Evidently the study of the
nonlinear stability of Bianchi models is very relevant to developing
this picture further.

Independently of the Fischer-Moncrief picture the study of small
(but finite) perturbations of Bianchi models is an avenue for 
making progress in understanding expanding cosmological models. 
There is a large literature on linear perturbations of 
cosmological models and it would be desirable to determine
what insights the results of this work might suggest for the
full nonlinear dynamics. Just as it is interesting to know under 
what circumstances homogeneous cosmological models become isotropic
in the course of expansion, it is interesting to know when more
general models become homogeneous. This does happen in the case
of small perturbations of the Milne model. On the other hand,
there is an apparent obstruction in other cases. This is the 
Jeans instability \cite{longair98a, boerner93a}.  A linear analysis 
indicates that
under certain circumstances (e.g. perturbations of a flat Friedmann
model) inhomogeneities grow with time. As yet there are no results
on this available for the fully nonlinear case. A comparison which
should be useful is that with Landau damping in plasma physics,
where rigorous results are available \cite{guo95a}.

The most popular matter model for spatially homogeneous cosmological
models is the perfect fluid. Generalizing this to inhomogeneous models
is problematic since formation of shocks or (in the case of dust) 
shell-crossing must be expected to occur. These signal an end to
the interval of evolution of the cosmological model which can be
treated mathematically with known techniques. Criteria for the
development of shocks (or their absence) should be developed, based
on the techniques of classical hydrodynamics. 

In the case of polarized Gowdy spacetimes there is a description of 
the late-time asymptotics in the literature \cite{chrusciel90b}, although 
the proofs have unfortunately never been published. The central object
in the analysis of these spacetimes is a function $P$ which satisfies
the equation $P_{tt}+t^{-1}P_t=P_{\theta\theta}$. The picture which 
emerges is that the leading asymptotics is given by $P=A\log t+B$ for
constants $A$ and $B$, this being the form taken by this function
in a general Kasner model, while the next order correction consists of 
waves whose amplitude decays like $t^{-1/2}$, where $t$ is the usual 
Gowdy time coordinate. The entire spacetime can be reconstructed from 
$P$ by integration. It turns out that the generalized Kasner exponents
converge to $(1,0,0)$ for inhomogeneous models. This shows that if 
it is stated that these models are approximated by Kasner models at
late times it is necessary to be careful in what sense the approximation 
is suposed to hold. Information
on the asymptotics is also available in the case of small but finite 
perturbations of the Milne model and the Bianchi type III form of flat
spacetime, as discussed in sections \ref{milne} and \ref{bianchi3}
respectively. 

There are not too many results on future geodesic completeness for
inhomogeneous cosmological models. A general criterion for geodesic
completeness is given in \cite{choquet02a}. It does not apply to
cases like the Kasner solution but is well-suited to the case where
the second fundamental form is eventually negative definite.

\subsection{Inflationary models}

One important aspect of the fragmentary picture of the dynamics of
expanding cosmological models presented in the last two sections is
that it seems to be complicated. A situation where we can hope for
a simpler, more unified picture is that where a positive cosmological
constant is present. Recall first that when the cosmological constant
vanishes and the matter satisfies the usual energy conditions spacetimes
of Bianchi type IX recollapse \cite{lin90a} and so never belong to the
indefinitely expanding models. When $\Lambda>0$ this is no longer 
true. Then Bianchi IX spacetimes show complicated features which
will not be considered here (cf. \cite{oliveira02a}). In discussing 
homogeneous models we
restrict to the other Bianchi types. Then a general theorem of
Wald \cite{wald83} states that any model whose matter content satisfies 
the strong and dominant energy conditions and which expands for an
infinite proper time $t$ is such that all generalized Kasner exponents
tend to $1/3$ as $t\to\infty$. A positive cosmological constant leads
to isotropization. The mean curvature tends to the constant value 
$-\sqrt{3\Lambda}$ as $t\to\infty$ while the scale factors increase
exponentially.

Wald's result is only dependent on energy conditions and uses no details
of the matter field equations. The question remains whether solutions
corresponding to initial data for the Einstein equations with positive
cosmological constant coupled to reasonable matter exist globally in 
time under the sole condition that the model is originally expanding. 
It can be shown that this is true for various matter models using the
techniques of \cite{rendall95c}. Suppose we have a solution on an 
interval $(t_1,t_2)$. It follows from \cite{wald83} that the mean 
curvature is increasing and no greater than $-\sqrt{3\Lambda}$. 
Hence, in particular, $\tr k$ is bounded as $t$ approaches $t_2$. 
Now we wish to verify condition (7) of \cite{rendall95c}
which says that if the mean curvature is bounded as an endpoint of the
interval where it is defined is approached then the solution can be
extended to a longer interval. As in \cite{rendall95c} it can be 
shown that if $\tr k$ is bounded then $g_{ij}$, $k_{ij}$ and $(\det g)^{-1}$
are bounded. Thus, in the terminology of \cite{rendall95c}, it is enough to
check (7)${}'$ for a given matter model in order to get the desired global
existence theorem. This condition involves the behaviour of a fluid in
a given spacetime. Since the Euler equation does not contain $\Lambda$
the result of \cite{rendall95c} applies directly. It follows that global
existence holds for perfect fluids and mixtures of non-interacting perfect
fluids. A similar result holds when the matter is described by collisionless
matter satisfying the Vlasov equation. Here it suffices to note that the
proof of Lemma 2.2 of \cite{rendall94b} generalizes without difficulty to
the case where a cosmological constant is present.   

The effect of a cosmological constant can be mimicked by a suitable
exotic matter field which violates the strong energy condition, for
example a nonlinear scalar field with exponential potential. In the
latter case an analogue of Wald's theorem has been proved by Kitada
and Maeda \cite{kitada92}. For a potential of the form $e^{-\lambda\phi}$ 
with $\lambda$ smaller than a certain limiting value the qualitative
picture is similar to that in the case of a positive cosmological 
constant. The difference is that the asymptotic rate of decay of certain
quantities is not the same as in the case with positive $\Lambda$. 
In \cite{kitada93} it is discussed how the limiting value of $\lambda$
can be increased. The behaviour of homogeneous and isotropic
models with general $\lambda$ has been investigated in \cite{halliwell87}.
 
Both models with a positive cosmological constant and models with
a scalar field with exponential potential are called
inflationary because the rate of (volume) expansion is increasing with
time. There is also another kind of inflationary behaviour which arises 
in the presence of a scalar field with power law potential like $\phi^4$ 
or $\phi^2$. In that case the inflationary property concerns the behaviour of
the model at intermediate times rather than at late times. The picture is
that at late times the universe resembles a dust model without cosmological
constant. This is known as reheating. The
dynamics has been analysed heuristically by Belinskii et. al. 
\cite{belinskii86a}. Part of their conclusions have been proved rigorously
in \cite{rendall01b}. Calculations analogous to those leading to a proof
of isotropization in the case of a positive cosmological constant or an 
exponential
potential have been done for a power law potential in \cite{moss86}. In 
that case the conclusion cannot apply to late time behaviour. Instead some 
estimates are obtained for the expansion rate at intermediate times.

Consider what happens to Wald's proof in an inhomogeneous spacetime with
positive cosmological constant. His arguments only use the Hamiltonian
constraint and the evolution equation for the mean curvature. In Gauss
coordinates spatial derivatives of the metric only enter these equations 
via the spatial scalar curvature in the Hamiltonian constraint. Hence,
as noticed in \cite{jensen87}, Wald's argument applies to the inhomogeneous
case, provided we have a spacetime which exists globally in the future in
Gauss coordinates and which has everywhere non-positive spatial scalar 
curvature. Unfortunately it is hard to see how the latter condition can
be verified starting from initial data. It not clear whether there is
a non-empty set of inhomogeneous initial data to which this argument
can be applied. 

In the vacuum case with positive cosmological constant the result of
Friedrich discussed in section \ref{desitter} proves local homogenization
of inhomogeneous spacetimes, i.e. that all generalized Kasner exponents
corresponding to a suitable spacelike foliation tend to $1/3$ in the 
limit. To see this, consider (part of) the de Sitter metric in the form
$-dt^2+e^{2t}(dx^2+dy^2+dz^2)$. This choice, which is different from
that discussed in \cite{friedrich86}, simplifies the algebra as much
as possible. Letting $\tau=e^{-t}$ shows that the above metric can be 
written in the form $\tau^{-2}(-d\tau^2+dx^2+dy^2+dz^2)$. This exhibits
the de Sitter metric as being conformal to a flat metric. In the 
construction of Friedrich the conformal class and conformal factor are
perturbed. The corrections to the metric in terms of coordinate 
components are of relative order $\tau=e^{-t}$. Thus the trace-free
part of the second fundamental forms decays exponentially, as desired.

There have been several numerical studies of inflation in inhomogeneous 
spacetimes. These are surveyed in Section 3 of \cite{anninos01a}.

\section{Structure of general singularities}\label{sing}

The aim of this section is to present a picture of the nature of
singularities in general solutions of the Einstein equations. It
is inspired by the ideas of Belinskii, Khalatnikov and Lifshitz 
(BKL). To fix ideas, consider the case of a solution of the
Einstein equations representing a cosmological model with a big
bang singularity. A central idea of the BKL picture is that
near the singularity the evolution at different spatial points 
decouples. This means that the global spatial topology of the
model plays no role. The decoupled equations are ordinary 
differential equations. They coincide with the equations for 
spatially homogeneous cosmological models, so that the study
of the latter is of particular significance.

\subsection{Lessons from homogeneous solutions}\label{homsing}

In the BKL picture a Gaussian coordinate system $(t,x^a)$ is introduced
such that the big bang singularity lies at $t=0$. It is not a priori
clear whether this should be possible for very general spacetimes. 
A positive indication is given by the results of \cite{andersson01a},
where coordinates of this type are introduced in one very general
class of spacetimes. Once these coordinates have been introduced
the BKL picture says that the solution of the Einstein equations
should be approximated near the singularity by a family of spatially
homogeneous solutions depending on the coordinates $x^a$ as parameters.
The spatially homogeneous solutions satisfy ordinary differential
equations in $t$. 

Spatially homogeneous solutions can be classified into Bianchi and
Kantowski-Sachs solutions. The Bianchi solutions in turn can be
subdivided into types I to IX according to the Lie algebra of the
isometry group of the spacetime. Two of the types, VI${}_h$ and
VII${}_h$ are in fact one-parameter families of non-isomorphic
Lie algebras labelled by $h$. The generality of the different 
symmetry types can be judged by counting the number of parameters
in the initial data for each type. The result of this is that the 
most general types are Bianchi VIII, Bianchi IX and Bianchi 
VI${}_{-1/9}$. The usual picture is that Bianchi VIII and Bianchi
IX have more complicated dynamics than all other types and that
the dynamics is similar in both these cases. This leads one to
concentrate on Bianchi type IX and the mixmaster solution (see
section \ref{homogeneous}). Bianchi type VI${}_{-1/9}$ was
apparently never mentioned in the work of BKL and has been 
largely ignored in the literature. This is a gap in understanding
that should be filled. Here we follow the majority and focus on
Bianchi type IX.

Another aspect of the BKL picture is that most types of matter should
become negligible near the singularity for suitably general solutions.
In the case of perfect fluid solutions of Bianchi type IX with a
linear equation of state this has been proved by Ringstr\"om
\cite{ringstrom00b}. In the case of collisionless matter it remains
an open issue, since rigorous results are confined to Bianchi types
I, II and III and Kantowski-Sachs and have nothing to say about 
Bianchi type IX. If it is accepted that matter is usually asymptotically
negligible then vacuum solutions become crucial. The vacuum solutions
of Bianchi type IX (mixmaster solutions) play a central role. They 
exhibit complicated oscillatory behaviour, and essential aspects of this
have been captured rigorously in the work of Ringstr\"om \cite{ringstrom00a,
ringstrom00b} (compare section \ref{homogeneous}). 

Some matter fields can have an important effect on the dynamics near
the singularity. A scalar field or stiff fluid leads to the
oscillatory behaviour being replaced by monotone behaviour of the
basic quantities near the singularity and thus to a great simplification
of the dynamics. An electromagnetic field can cause oscillatory
behaviour which is not present in vacuum models or models with perfect
fluid of the same symmetry type. For instance models of Bianchi type I
with an electromagnetic field show oscillatory, mixmaster-like
behaviour \cite{leblanc97a}. However it seems that this does not lead
to anything essentially new. It is simply that the effects of spatial
curvature in the more complicated Bianchi types can be replaced by 
electromagnetic fields in simpler Bianchi types.

A useful heuristic picture which systematizes much of what is known 
about the qualitative dynamical behaviour of spatially homogeneous
solutions of the Einstein equations is the idea developed by Misner 
\cite{misner67a} of representing the dynamics as the motion of a particle
in a time-dependent potential. In the approach to the singularity
the potential develops steep walls where the particle is reflected.
The mixmaster evolution consists of an infinite sequence of bounces
of this kind. 

\subsection{Inhomogeneous solutions}\label{inhomsing}

Consider now inhomogeneous solutions of the Einstein equations where,
according to the BKL picture, oscillations of mixmaster type are to be
expected. This is for instance the case for general solutions of the
vacuum Einstein equations. There is only one rigorous result
to confirm the presence of these oscillations in an inhomogeneous
spacetime of any type and that concerns a family of spacetimes depending
on only finitely many parameters \cite{berger00a}. They are obtained by 
applying a solution-generating technique to the mixmaster solution.
Perhaps a reason for the dearth of results is that oscillations
usually only occur in combination with the formation of local spatial
structure discussed in section \ref{locstruc}. On the other hand there 
is a rich variety of numerical and heuristic work supporting the
BKL picture in the inhomogeneous case \cite{berger02a}.

A situation where there is more hope of obtaining rigorous results
is where the BKL picture suggests that there should be monotone
behaviour near the singularity. This is the situation where Fuchsian
techniques can often be applied to prove the existence of large classes
of spacetimes having the expected behaviour near the initial
singularity (see section \ref{fuchsian}). It would be desirable to
have a stronger statement than these techniques have provided up
to now. Ideally it should be shown that a non-empty open set of solutions
of the given class (by which is meant all solutions corresponding
to an open set of initial data on a regular Cauchy surface) lead to
a singularity of the given type. The only results of this type in 
the literature concern polarized Gowdy spacetimes \cite{isenberg90}, 
plane symmetric spacetimes with a massless scalar field \cite{rendall95b},
spacetimes with collisionless matter and spherical, plane or
hyperbolic symmetry \cite{rein96a} and a subset of general Gowdy 
spacetimes \cite{chrusciel91a}. The work of Christodoulou 
\cite{christodoulou87b} on spherically symmetric solutions of the 
Einstein equations with a massless scalar field should also be mentioned 
in this context although it concerns the singularity inside a black hole 
rather than singularities in cosmological models. Note that all these
spacetimes have at least two Killing vectors so that the PDE problem
to be solved reduces to an effective problem in one space dimension.

\subsection{Formation of localized structure}\label{locstruc}

Numerical calculations and heuristic methods such as those used by BKL
lead to the conclusion that as the singularity is approached localized
spatial structure will be formed. At any given spatial point the
dynamics is approximated by that of a spatially homogeneous model
near the singularity and there will in general be bounces (cf.
section \ref{homsing}). However there will be exceptional spatial
points where the bounce fails to happen. This leads to a situation
where the spatial derivatives of the quantities describing the
geometry blow up faster than these quantities themselves as the
singularity is approached. In general spacetimes there will be
infinitely many bounces before the singularity is reached and so 
the points where the spatial derivatives are large will get more 
and more closely separated as the singularity is approached.

In Gowdy spacetimes only a finite number of bounces are to be expected
and the behaviour is eventually monotone (no more bounces).
There is only one essential spatial dimension due to the symmetry and 
so large derivatives in general occur at isolated values of the
one interesting spatial coordinate. Of course these correspond to 
surfaces in space when the symmetry directions are restored. The
existence of Gowdy solutions showing features of this kind has been
proved in \cite{rendall01a}. This was done by means of an explicit
transformation which makes use of the symmetry. Techniques should
be developed which can handle this type of phenomenon more directly
and more generally.

The formation of spatial structure calls the BKL picture into question
(cf. the remarks in \cite{belinskii92a}). The basic assumption underlying
the BKL analysis is that spatial derivatives do not become too large
near the singularity. Following the argument to its logical conclusion
then indicates that spatial derivatives do become large near a dense
set of points on the initial singularity. Given that the BKL picture
has given so many correct insights, the hope that it may be generally
applicable should not be abandoned too quickly. However the problem
represented by the formation of spatial structure shows that at the 
very least it is necessary to think carefully about the sense in which
the BKL picture could provide a good approximation to the structure of
general spacetime singularities. 

\section{Further results}\label{further}

This section collects miscellaneous results which do not fit into the main
line of the exposition.

\subsection{Evolution of hyperboloidal data}\label{hyperboloidal}

In section~\ref{constraints} hyperboloidal initial data were mentioned. They 
can be thought 
of as generalizations of the data induced by Minkowski space on a hyperboloid.
In the case of Minkowski space the solution admits a conformal 
compactification where a conformal boundary, null infinity, can be added to
the spacetime. It can be shown that in the case of the maximal development
of hyperboloidal data a piece of null infinity can be attached to the 
spacetime. For small data, i.~e.\ data close to that of a hyperboloid in
Minkowski space, this conformal boundary also has completeness properties
in the future allowing an additional point $i_+$ to be attached there. (See
~\cite{friedrich91} and references therein for more details.) Making contact 
between hyperboloidal data and asymptotically flat initial data is much more 
difficult and there is as yet no complete picture. (An account of the
results obtained up to now is given in~\cite{friedrich98a}.) If the relation
between hyperboloidal and asymptotically flat initial data could be
understood it would give a very different approach to the problem
treated by Christodoulou and Klainerman (section~\ref{minkowski}).
It might well also give more detailed information on the asymptotic
behaviour of the solutions. 

\subsection{The Newtonian limit}\label{limit}

Most textbooks on general relativity discuss the fact that Newtonian
gravitational theory is the limit of general relativity as the speed
of light tends to infinity. It is a non-trivial task to give a precise
mathematical formulation of this statement. Ehlers systematized extensive 
earlier work on this problem and gave a precise definition of the Newtonian 
limit of general relativity which encodes those properties which are 
desirable on physical grounds (see~\cite{ehlers91}.) Once a definition has 
been given the question remains whether this definition is compatible with
the Einstein equations in the sense that there are general families 
of solutions of the Einstein equations which have a Newtonian limit
in the sense of the chosen definition. A theorem of this kind was
proved in~\cite{rendall94a}, where the matter content of spacetime was assumed 
to be a collisionless gas described by the Vlasov equation. (For another
suggestion as to how this problem could be approached see 
~\cite{fritelli94}.) The essential mathematical problem is that of a
family of equations depending continuously on a parameter
$\lambda$ which are hyperbolic for $\lambda\ne 0$ and degenerate for
$\lambda=0$. Because of the singular nature of the limit it is by no
means clear a priori that there are families of solutions which depend 
continuously on $\lambda$. That there is an abundant supply of families
of this kind is the result of~\cite{rendall94a}. Asking whether there are 
families which are $k$ times continuously differentiable in their dependence 
on $\lambda$ is related to the issue of giving a mathematical justification of
post-Newtonian approximations. The approach of~\cite{rendall94a} has not even 
been extended to the case $k=1$ and it would be desirable to do this. Note
however that for $k$ too large serious restrictions arise~\cite{rendall92a}. 
The latter fact corresponds to the well-known divergent behaviour of higher 
order post-Newtonian approximations.  

It may be useful for practical projects, for instance those based on 
numerical calculations, to use hybrid models where the equations for 
self-gravitating Newtonian matter are modified by terms representing 
radiation damping. If we expand in terms of the parameter $\lambda$ as 
above then at some stage radiation damping 
terms should play a role. The hybrid models are obtained by truncating these
expansions in a certain way. The kind of expansion which has just been 
mentioned can also be done, at least formally, in the case of the Maxwell
equations. In that case a theorem on global existence and asymptotic 
behaviour for one of the hybrid models has been proved in \cite{kunze01a}.
These results have been put into context and related to the Newtonian limit
of the Einstein equations in \cite{kunze01b}. 

\subsection{Newtonian cosmology}\label{cosmology}

Apart from the interest of the Newtonian limit, Newtonian gravitational
theory itself may provide interesting lessons for general relativity.
This is no less true for existence theorems than for other issues.
In this context it is also interesting to consider a slight generalization
of Newtonian theory, the Newton-Cartan theory. This allows a nice
treatment of cosmological models, which are in conflict with the (sometimes
implicit) assumption in Newtonian gravitational theory that only isolated 
systems are considered. It is also unproblematic to introduce a 
cosmological constant into the Newton-Cartan theory.

Three global existence theorems have been proved in Newtonian cosmology.
The first~\cite{brauer94} is an analogue of the cosmic no hair theorem 
(cf.\ section~\ref{desitter}) and concerns models with a positive cosmological 
constant. It asserts that homogeneous and isotropic models are nonlinearly 
stable if 
the matter is described by dust or a polytropic fluid with pressure. Thus it 
gives information about global existence and asymptotic behaviour for models
arising from small (but finite) perturbations of homogeneous and isotropic
data. The second and third results concern collisionless matter and the
case of vanishing cosmological constant. The second~\cite{rein94b} says that 
data which constitute a periodic (but not necessarily small) perturbation of
a homogeneous and isotropic model which expands indefinitely give rise to
solutions which exist globally in the future. The third~\cite{rein97} says 
that the homogeneous and isotropic models in Newtonian cosmology which 
correspond to a $k=-1$ Friedmann-Robertson-Walker model in general relativity 
are non-linearly stable.

\subsection{The characteristic initial value problem}\label{char}

In the standard Cauchy problem, which has been the basic set-up for all the
previous sections, initial data are given on a spacelike hypersurface.
However there is also another possibility, where data are given on one or 
more null hypersurfaces. This is the characteristic initial value problem.
It has the advantage over the Cauchy problem that the constraints reduce
to ordinary differential equations. One variant is to give initial data
on two smooth null hypersurfaces which intersect transversely in a
spacelike surface. A local existence theorem for the Einstein equations 
with an initial configuration of this type was proved in~\cite{rendall90}.
Another variant is to give data on a light cone. In that case local existence
for the Einstein equations has not been proved, although it has been proved 
for a class of quasilinear hyperbolic equations which includes the reduced 
Einstein equations in harmonic coordinates~\cite{Dossa97}. 

Another existence theorem which does not use the standard Cauchy problem,
and which is closely connected to the use of null hypersurfaces, concerns
the Robinson-Trautman solutions of the vacuum Einstein equations. In that
case the Einstein equations reduce to a parabolic equation. Global existence
for this equation has been proved by Chru\'sciel~\cite{Chrusciel91b}.

\subsection{The initial boundary value problem}\label{ibvp}

In most applications of evolution equations in physics (and in other 
sciences) initial conditions need to be supplemented by boundary conditions.
This leads to the consideration of initial boundary value problems. It
is not so natural to consider such problems in the case of the Einstein 
equations since in that case there are no physically motivated boundary
conditions. (For instance, we do not know how to build a mirror for 
gravitational waves.) An exception is the case of a fluid boundary discussed
in section~\ref{freeboundary}.

For the vacuum Einstein equations it is not a priori clear that it is even
possible to find a well-posed initial boundary value problem. Thus it is
particularly interesting that Friedrich and Nagy~\cite{friedrich99a} have 
been able to prove the well-posedness of certain initial boundary value
problems for the vacuum Einstein equations. Since boundary conditions 
come up quite naturally when the Einstein equations are solved numerically,
due to the need to use a finite grid, the results of~\cite{friedrich99a}
are potentially important for numerical relativity. The techniques 
developed there could also play a key role in the study of the initial
value problem for fluid bodies (Cf.\ section~\ref{freeboundary}.) 
 
\section{Acknowledgements}
I thank H\aa kan Andr{\'e}asson, Bernd Br{\"u}gmann, John Wainwright and 
Marsha Weaver for helpful suggestions.

\bibliography{existence7}

\end{document}